# Impurity-induced Polar States in SrTiO$_3$ Quantum Paraelectric

P.A. Markovin[1,4], V.A. Trepakov[1,2], M.E. Guzhva[1,4,6], A.G. Razdobarin[1],

A.K. Tagantsev [1,3], D.A. Andreev[4], M. Itoh[5] and A. Dejneka[2]

[1] Ioffe Physical-Technical Institute of the RAS, 194021, Saint-Petersburg, Russia

[2] Institute of Physics, ASCR, Na Slovance 2, 182 21 Praha 8, Czech Republic

[3] Swiss Federal Institute of Technology (EPFL), Lausanne, 1015, Switzerland

[4] Saint-Petersburg State Polytechnical University, 195251, Saint-Petersburg, Russia

[5] Tokyo Institute of Technology, 4259 Nagatuta, Midori, Yokohama 226-8503, Japan

[6] National Research University for Information Technology, Mechanics and Optics, 197101

Saint-Petersburg, Russia

E-mail: p.markovin@mail.ioffe.ru

**Abstract**. Short and long range impurity-induced polar ordering in Sr$_{1-x}$Ca$_x$TiO$_3$ with x = 0.014 (SCT-1.4) and SrTi($^{16}$O$_{0.03}$$^{18}$O$_{0.97}$)$_3$ (STO-18) single crystals were investigated and discussed on the basis of the light refraction temperature dependences measurements. For SCT-1.4 the temperature dependences of the morphic birefringence and dielectric hysteresis loops were measured. Within the optical indicatrix perturbation approach a method for calculation from polar contributions to the refraction of light of short-range polarization contribution P$_{sh}$ originated by spatial fluctuations is developed for the systems with the coexistence of long and short range polar ordering. The magnitudes and temperature dependences of P$_{sh}$ and long-range spontaneous polarization P$_s$ have been determined in SCT-1.4 and STO-18. The results allowed to characterize quantitatively and to compare contributions of short (P$_{sh}$) and long (P$_s$) range ordering in the formation of impurity-induced polar phase in SrTiO$_3$.


# 1. INTRODUCTION

The well-known quantum paraelectrics $SrTiO_3$ (STO) and $KTaO_3$ (KTO) are very sensitive to doping with impurities and defects. A low-temperature (low-T) polar state can be easily induced in these materials by incorporating a sufficient quantity of suitable "ferroelectric active" impurities above a critical concentration $x_c$ ([1-10] and references therein). These impurities can be considered as: *i*) off-center substitution ions with reorientable electric dipole moments, e.g., $Li^+$ in KTO, in which the coupling between the $TO_1$ soft polar mode and the mutual dipole-dipole interaction is mediated by a $TO_1$ mode, thereby forming a dipole glass or a ferroelectric (FE) state; *ii*) on-center substitution ions that are larger than or nearly the same size as the substituted host ions in, among others, $Sr_{1-x}Ba_xTiO_3$ (SBT), $Sr_{1-x}Ca_xTiO_3$ (SCT), $KTa_{1-x}Nb_xO_3$ (KNT); and *iii*) the oxygen isotope exchange system $SrTi(^{16}O_{1-x}{}^{18}O_x)_3$ (STO-18). The characteristics, structure and formation of the low-T polar phase remain controversial in both theory and experiments, even for the well-documented SCT, STO-18 and KTN phases. According to a widely used concept, ferroelectric behavior of these systems in the most cases is connected to off-central position of impurity ions [1-6]. Vikhnin, Markovin, Lemanov and Kleemann [11] suggested that the low-T phase transformations in doped incipient ferroelectrics involve presence of compositional clusters.

The interaction of these clusters with the $TO_1$ lattice mode leads to a displacive-type FE phase transition (PT). When the clusters have reorientable dipole moments, the order-disorder FE or dipole glass-like PT occurs within a percolation scenario. In [7,8] (Kvyatkovskii), the critical concentration $x_c$ for KTN and $Sr_{1-x}A_xTiO_3$ solid solutions was calculated using a virtual crystal approach in soft FE mode theory. The resulting $x_c$ values agreed well with the available experimental data. Kvyatkovskii formulated a theory for STO-18 in [7, 12] in which the FE lattice instability was considered to be an anharmonic and quantum effect. The associated FE PT was described in terms of the suppression of the quantum mechanical zero-point atomic motion because of the enhancement of the atomic masses by substituting a heavy isotope for a light isotope. Bussmann-Holder, Büttner and Bishop [13,14] modeled an isotope-induced FE PT using a nonlinear electron-phonon interaction model that was based on the configurational instability of the oxygen ions and, consequently, dynamical *p-d* hybridization effects. The PT is first order and is likely to be accompanied by precursor structures, such as twinning and tweed structures [13]. Far above the PT point, polar micro-domains are formed that increase in size at cooling and freeze out at $T_c$ without long range ordering. Displacive and order-disorder features coexist with each other. Both components persist in the FE phase, forming an incomplete and inhomogeneous FE low-T state



[14]. STO-18 is currently a highly active subject in experimental studies. Soft mode dynamics have been clearly identified in the hyper-Raman range [15], in the "perfect softening" of the Slater $E_u$ FE mode in ultra-low-frequency Raman experiments [16]. The dielectric permittivity measurements under hydrostatic pressure and in $dc$ electric field were studied [17] and spontaneous polarization measurements from dielectric hysteresis loops [10, 18, 19] were done. NMR and EPR experiments have provided evidence of Ti disorder and that an order-disorder component exists in addition to the displacive soft mode component [20-22]. The FE ordering develops in two stages. Below 70 K, rhombohedral clusters are formed in the $D^{18}_{4h}$ tetragonal host matrix. Upon further cooling, these clusters increase in size and concentration, freeze out, and percolate, resulting in an inhomogeneous FE state below $T_c = 24$ K. Remarkably, Ti disorder, a two-component state and similar clusters were also reported in [20] for the conventional quantum paraelectric STO, but the cluster population was too small to support percolation. Raman experiments on STO-18 with $x$ values of 0.23 and 0.32 that were below the critical value $x_c \approx 0.33$ [23] showed that crystal inhomogeneity was enhanced by the exchange of $^{18}$O, which plays an essential role in the soft-mode dynamics of STO-18 [24]. Dielectric and nonlinear susceptibility studies in [25-29] showed the absence of long-range order in the low-T phase, the presence of polar nanoregions and random fields and the importance of disorder. Studies combining optical second harmonic measurements with neutron and X-ray scattering [30] have shown that long-range order in the polar phase does not develop over the entire crystal but is frozen into local regions. However, in Raman experiments, the ideal soft-mode type quantum PT has been reported to depend on $x$ [31]. Thus, it has been concluded that $^{18}$O exchange enhances $TO_1$ mode softening by suppressing quantum fluctuations. In the vicinity of the critical point ($x_c \approx 0.33$), the system exhibits FE-paraelectric phase coexistence, whereas a homogeneous FE phase forms for STO-18-$x$ when $x$ is sufficiently larger than $x_c$. The mixed systems, SCT and KTN, have attracted considerable attention for many years (see, for example [3-5, 32-36, **57**] and the references therein). So, as it is seen from analysis given above, in spite of decades of investigations, polar ordering phenomena in quantum paraelectrics induced by impurities are continuing to be of high interest and controversies and there is no general agreement about its nature and character.

Unusual properties of STO and KTO crystals doped with "FE active" impurities and properties of STO-18 with $x \geq x_c$ have been reported in number experimental studies. It was shown, that polar short range ordering emerges at temperatures appreciably above $T_C$, and below $T_C$ (and at high enough x > $x_c$) the low-T polar state reveals characteristic manifestations of both the long- and short-range ordered phases. For interpretation of such state different speculative models and the



corresponding terms have been used (ferroelectric glass – coexistence of ferroelectricity and dipole glass, local ordering, nanodomains, ferro-phase with fluctuations e. t. c. ([1-5, 20, 35-37, 57]).

The general problem for the systems with coexistence of long and short ordering, in particular for ones with polar ordering, is the absence of correct methods of experiments treatment for quantitative determination of short range order polarization contribution $P_{sh}$ below $T_c$ and as a result the absence of the quantitative magnitude of $P_{sh}$. On the other hand the magnitude $P_{sh}$ as well as the spontaneous polarization $P_s$, which describes long polar ordering, should be the direct and quantitative characteristic of short range ordering contributions into polar phase formation and important parameter for theoretical descriptions. It is clear that $P_{sh}$ as a quantitative parameter for the description only short range order have to be obtained as the root mean square of the short range order polarization $\left\langle P_{sh}^2 \right\rangle^{1/2}$ since $\left\langle P_{sh} \right\rangle = 0$.

Up to now experientially from the temperature dependencies of the refractive index and principal birefringence the magnitudes of $P_{sh} = \left\langle P_{sh}^2 \right\rangle^{1/2}$ were derived separately only in special simple cases when the long range spontaneous polarization $P_s = 0$. In the case of polar ordering the spontaneous polar contribution to the refractive index is proportional via electro-optical coefficients to the square of the spontaneous polarization $P^2$ [38]. The magnitude of $P$, obtained by this method, really related to the root mean square of the polarization $\left\langle P^2 \right\rangle^{1/2}$, due to time and space-averaged of the square of the polarization $P^2$ over ordering on a long and short length scale, so $\left\langle P^2 \right\rangle^{1/2} = P_{sh}$ when long range order is absent. The example is the local ordering in polar nanoregions of Burns phase in relaxors [39-42]. $P_d$ in polar nanoregions ($P_{sh}$ in our notation) evaluating as $\left\langle P^2 \right\rangle^{1/2}$ from the polar contribution to the temperature dependence of refractive index arise sufficiently above the temperature of the dielectric permittivity maxima without ordering on a long length scale. Another ones are the determination from refraction experiments the magnitude and the temperature dependence of $P_{sh}$ as the precursor polarization (fluctuations of the order parameter - short range order) above $T_C$ in classical ferroelectrics $BaTiO_3$ [43], $KNbO_3$ [44], $PbTiO_3$ [45] and in $TlTiOPO_4$ crystals [46].

In quantum paraelectrics with impurities as it is noted above the low-T polar state reveals characteristic manifestations of both the long- and short-range ordered phases below $T_C$ (even at high enough $x > x_c$). Taking into account the actuality of refractive index and birefringence studies



of short range order at the end of introduction the findings of [35, 36, 47, 48, 57] should be highlighted. The magnitude of the short-range order polarization $P_{sh} = \left\langle P^2 \right\rangle^{1/2}$ have been determined in $Sr_{1-x}Ba_xTiO_3$ from the refraction of light [47, 48] at x = 0.02, where long range is absent. At higher Ba concentration the magnitude of $\left\langle P^2 \right\rangle^{1/2}$ include short- and long-range order. In [36] the polarization $\left\langle P^2 \right\rangle^{1/2}$, related to the both, short- and long range order, in $KTa_{1-x}Nb_xO_3$ (x=0.008, 0.012, 0.02) was determined from the temperature dependence of the polar contribution to the refractive index and appeared to be considerably larger than the polarization, obtained from morphic birefringence measurements. Obtained result qualitatively characterizes large contribution of short ordering in polar phase formation below $T_C$. However, magnitude of $\left\langle P^2 \right\rangle^{1/2}$ is not a correct enough as quantitative characteristic of polarization for regions with coexistence of short range order below $T_C$. It includes contributions as long as short order and contribution of the last one cannot be obtained by simple deduction $\left\langle P^2 \right\rangle^{1/2}$ - $P_s$. In [35, 57] temperature dependence of principal birefringence and morphic birefringence were measured in SCT (x=0.007) and principal birefringence in SCT (x=0.002). It was shown, that the polar contribution to the principal birefringence in SCT (x=0.007) was considerably larger than morphic birefringence. The $P_s$ magnitude via $\left\langle P^2 \right\rangle^{1/2}$ has been estimated. The results allow authors of [57] to suggest that SCT (x=0.007) exhibits an inhomogeneous induced polar phase with ordering at two length scales: local ordering associated with individual nanodomains, ordering on a longer length scale due to large correlated domains. Temperature dependences of the principal birefringence and morphic birefringence for STO-18 were measured in [37, 49]. Qualitatively the results show the presence long and short-range orders. The magnitudes of $\left\langle P^2 \right\rangle^{1/2}$ and $P_s$ have not been evaluated.

To clarify quantitatively the role of short ($P_{sh}$) and long ($P_s$) range ordering in the formation of impurity-induced polar states in $Sr_{1-x}Ca_xTiO_3$ with x = 0.014 (SCT-1.4) and $SrTi(^{16}O_{0.03}{}^{18}O_{0.97})_3$ (STO-18) the temperature dependence of the principal values of the refraction (specific optical retardation) of light in STO-18, SCT-1.4 and nominally pure $SrTiO_3$ (STO) single crystals have been measured and the magnitudes and temperature variations of polar contributions to the refraction in STO-18 and SCT-1.4 were derived. A method for quantitative calculations separately of $P_{sh}$ and $P_s$ below $T_C$ from polar contributions to the refraction of light was developed within the



framework of phenomenological indicatrix perturbation theory in terms of a generalized expression of polarization variation (including short range order) of the dielectric impermeability components. Additionally the temperature dependences of the morphic birefringence, dielectric hysteresis loops for SCT-1.4 and electrooptic (EO) coefficients for SCT-1.4 and STO-18 were measured in order to calculate more carefully and strictly separately $P_{sh}$ and $P_s$ below $T_C$. For these purpose we used also morphic birefringence and hysteresis loops data (STO-18) reported earlier in [37] and [10] respectively with the participation of one of our co-author M. Itoh. Hysteresis loops data (STO-18) from [18, 19] were also discussed. As a result we determined for SCT-1.4 and STO-18 the magnitude and temperature dependence of $P_{sh}$ polarization which appears only due to strong spatial fluctuations ("local" or "short range" ordering) and leads to contribution of short-range order in the polar phase below and above $T_C$ as well as the magnitude and temperature dependence of long range order polarization $P_s$.

Thus the main objective of this report is the correct determination of quantitative characteristics of the long- and short-range order contributions to impurity induced polar state in STO.

## 2. EXPERIMENTAL

### 2.1. Samples and experimental techniques

The flame fusion technique was used to grow nominally pure $SrTiO_3$ and $Sr_{1-x}Ca_xTaO_3$ single crystals with x = 0.014 (denoted below as SCT-1.4). Electron microprobe analysis was used to determine the chemical composition and to check the uniform Ca distribution, which proved to be homogeneous with fluctuation $\Delta x$ of $\pm$ 0.0005 lying in the range of counting statistic. This concentration was also controlled by the value of the antiferrodistorsive $O_h^1 - D_{4h}^{18}$ phase transition temperature $T_a$ which was determined as the temperature of the change of slope of the specific optical retardation temperature dependence and appearance of the crystallographic (principal) birefringence in $\Delta n_{ac}$ (T) dependence (according to [50] $T_a$ depends linearly on Ca concentration). The synthesis of $SrTi(^{16}O^{18}O_{0.97})$ single crystals (STO-18) has been described in [10]. Rectangular experimental platelets were cut along the pseudo cubic $[110]_a$, $[1\bar{1}0]_b$, and $[001]_c$ crystalline axes and spatially extended $\approx$ 2.5 mm along the $[001]_c$ axis and 0.3 ÷ 0.8 mm along the $[1\bar{1}0]_b$ axis (Fig.1). The directions and notations correspond to the tetragonal axes (*a, b, c*) below the transition



temperature $T_a$. The magnitude of critical concentration for $Sr_{1-x}Ca_xTaO_3$ is $x_c = 0.0018$ [51] and for $SrTi(^{16}O_{1-x}{}^{18}O_x)$ is $x_c \approx 0.33$ [23].

The temperature dependences of the principle values of the refraction of light in STO-18, SCT-1.4, nominally pure $SrTiO_3$ (STO) single crystals and the electrooptic (EO) effect at room temperature (RT) in STO-18, SCT-1.4 were measured at the wavelength 632.8 nm of a He-Ne laser using two-beam homodyne interferometer [52, 53]. The interferometric technique allows us to measure variations of the optical phase difference or the specific optical retardation $\delta\Psi_i$, which is related to the variation of the principle values of refractive index $\delta n_i(T)$ (see below the expression 1). The sensitivity of this technique is $\delta\Psi_i \approx 10^{-6}$ and an absolute accuracy in prolonged temperature measurements during $\approx 1$ hour of about $10^{-5}$ at a crystal thickness $l \approx 1$ mm. The advantages of the method are its very low variation of sensitivity with respect to changes of the light intensity transmitted by the sample, as well as low sensitivity to the influence of the depolarization of light passed through the sample. The temperature dependences of the refraction of light and the EO effect at RT were used to find the spontaneous polar contribution to the refraction of light, related to both long and short range ordering, for postcalculations of $P_s$ and $P_{sh}$. In these measurements light beam propagates along $\left[1\,\bar{1}\,0\right]_b$ axis.

In SCT-1.4 the morphic spontaneous polar birefringence $\Delta n_{ab} = n_b - n_a$ were measured for independently postcalculations of $P_s$. In this case light beam propagates along $\left[001\right]_c$ axis. Birefringence was measured using the polarimeter [53] at the wavelength 632.8 nm of a He-Ne laser. The sensitivity of this technique is $\Delta n \approx 10^{-7}$ at an absolute accuracy of about $10^{-6}$ at a crystal thickness $l \approx 1$ mm. Morphic birefringence is sensitive to the ordering on a long length scale - long range ordering which manifests symmetry change on the macroscopic scale [35-38]. In the case of the isotropic fluctuations of the order parameter (i.e., the polarization) in the $ab$-plane, the contribution of short-range order $P_{sh}$ in $\Delta n_{ab} = n_b - n_a$ is absent. It follows for $Sr_{1-x}Ca_xTaO_3$ x = 0.007 (denoted below as SCT-0.7) and STO-18 from [35], [37] respectively and SCT-1.4 from our measurements (see below Fig. 4 and Fig. 5, there are no fluctuation tails in morphic birefringence above $T_C$.). It is important to note, that symmetry considerations for spontaneous polar contribution to the refraction of light are universal. In the case, when space fluctuations of the order parameter are absent, in the single domain samples from $\delta n_i(T)$ measurements and the morphic polar birefringence one obtains the same magnitude of the spontaneous polarization $P_s$.



Also for the determining independently $P_s$ in SCT-1.4 and SCT-0.7 and to compare with $P_s$ obtained from optical measurements a conventional dielectric hysteresis loops technique was used (the data of $P_s$ from hysteresis loops measurements for STO-18 were taken from [10, 18, 19]). A modified Sawyer–Tower circuit was developed to measure the dielectric hysteresis loops and correspondingly the temperature dependences of spontaneous polarization in $Sr_{1-x}Ca_xTiO_3$. The setup allowed us to measure objects with small spontaneous and residual polarizations within the frequency range from 50 Hz to 1 kHz. Dielectric hysteresis loops measured at *ac* frequency 50 Hz are shown in Fig. 6 and 7 (electric field is applied along $[1\bar{1}0]_b$). Measurements performed at 1 kHz reveal nearly the same results.

The temperature of samples mounted in a helium cryostat was monitored independently using copper–constantan, iron-doped copper–copper thermocouples and a KTG-type semiconductor pickup. During the experiment, the measurement procedure was controlled on a computer and, depending on the actual measurement regime, either a fixed sample temperature or rate of its variation was maintained. The absolute error of temperature measurement did not exceed 0.5 K, and the sensitivity was 0.1 K throughout the temperature range covered, 5–300 K for hysteresis loops measurements (samples are in helium atmosphere) and 13-300 K for optical measurements (samples are in vacuum).

## 2.2. Calculations of the spontaneous polarization from interferometric measurements

The interferometric technique allows us to measure variations of the optical phase difference or the specific optical retardation $\delta\Psi_i$, which is related to the variation of the principle values of refractive index $\delta n_i(T)$ and the linear sample dilation along the propagation direction,

$$\delta\Psi_i(T) = \delta n_i(T) - (n_i - 1)\frac{\delta l_j(T)}{l_j} \qquad (1)$$

where $l_j$ is the sample thickness along of light propagation, $\delta l_j$ - is the change in the aforementioned thickness under an external perturbation, $n_i$ – is the principal values of the refractive index, index $i$ relates to the polarization of light, and $j$ to the direction of its propagation in the crystal. The spontaneous polar contribution to the optical retardation can be written as

$$\delta\Psi_i^s = \delta n_i^s + (n_i - 1)\frac{\delta l_j^s}{l_j} \qquad (2)$$



where $\delta n_i^s$ denotes the spontaneous polar contribution to the principal values of the refractive index, $\delta l_j^s$ denotes the spontaneous polar contribution to the temperature dilatations.

Relationship between spontaneous polar contributions to the refractive index and spontaneous polarization is

$$\delta n_i^s = \left[ -\frac{n_i^3}{2} g_{i3}^* \right] \left\langle P_3^2 \right\rangle \tag{3}$$

Where $\delta n_i^s$ denotes the spontaneous polar contribution to the principal values of the refractive index, and the $g_{i3}^*$ are the quadratic EO coefficients in the polar $C_{2v}$ phase (the symmetry of low temperature polar phase in SCT and STO-18) that are related to the $g_{ij}$ coefficient in the cubic phase of the oxygen-octahedral ferroelectrics through the relations given below [38, 48, 54]. $\left\langle P_3^2 \right\rangle$ is the average square of the spontaneous polarization $P_3$, lying along the axis indexed as 3.

The second summand in (2) is the contribution of the spontaneous deformation along the direction of light propagation that is related to the value of $\left\langle P_3^2 \right\rangle$ over the electrostriction coefficients $Q_{ij}^*$ (quadratic electrostriction):

$$\frac{\delta l_j^s}{l_j} = Q_{j3}^* \left\langle P_3^2 \right\rangle \tag{4}$$

$$\delta \Psi_i^s = \left[ -\frac{n_i^3}{2} g_{i3}^* + \left( n_i - 1 \right) Q_{j3}^* \right] \left\langle P_3^2 \right\rangle \tag{5}$$

In interferometric studies on the quadratic EO effect in cubic central inversion paraelectric phase at RT the magnitude of $\delta\Psi_i(E_3)$ relates to the electric field $E_3$ as follows:

$$\delta \Psi_i (E_3) = \left[ -\frac{n_i^3}{2} g_{i3}^* + \left( n_i - 1 \right) Q_{j3}^* \right] \varepsilon_0^2 (\varepsilon_3 - 1)^2 E_3^2 \tag{6}$$

Where $\varepsilon_0 (\varepsilon_{33} - 1) E_3 = P_{ind}(E_3)$ is the electric field induced polarization, $\varepsilon_0$ is the dielectric permittivity of free space, $\varepsilon_3$ is the dielectric permittivity along axes 3.

The $Q_{ij}^*$ and $g_{ij}^*$ tensors have similar symmetry properties; therefore, the expression in the square brackets in (5), (6) can be considered to be "effective EO coefficients". Thus, these coefficients can be adapted to the FE $C_{2v}$ phase using formulas analogous to those given in Table 1. The



interferometric EO studies were used to measure these same effective EO coefficients, which were used in turn to determine $\left\langle P_3^2 \right\rangle$ from $\delta\Psi_i^s$.

Consequently, $\left\langle P_3^2 \right\rangle$ can be determined quantitatively and accurately from interferometric measurements of $\delta\Psi_i(T)$ (after extracting $\delta\Psi_i^s$) and $\delta\Psi_i(E)$ without measuring the thermal expansion changes and electrostriction. Thus using the expression "refraction of light" we include both contributions into $\delta\Psi_i(T)$ (from refractive index and temperature dilatations).

Table 1 EO coefficients $g_{ij}^*$ in the ferroelectric phase $C_{2v}$ from EO coefficients $g_{ij}$ in cubic phase at RT

| $g_{13}^*$ | $g_{23}^*$ | $g_{33}^*$ |
|---|---|---|
| $g_{12}$ | $(1/2)(g_{11} + g_{12} - g_{44})$ | $(1/2)(g_{11} + g_{12} + g_{44})$ |

## 3. RESULTS AND DISCUSSIONS

From results of our measurements follows that quadratic EO coefficients $g_{ij}$ at RT in SCT-1.4 and STO-18 crystals is appeared to have almost of the same magnitude as that in nominally pure STO [54], which were measured also by interferrometric method. To calculate polarizations from the spontaneous contribution $\delta\Psi_i^s$ for SCT-1.4 and STO-18 from the optical measurements, the following values were used: $g_{11} = 0.15 m^4/C^2$, $g_{12} = 0.04 m^4/C^2$, $g_{44} = 0.08 m^4/C^2$ and a refractive index $n = 2.39$. The quadratic EO coefficients of STO were shown to be temperature independent within $4 - 300$ K temperature range [55]. The same behavior was also assumed for SCT-1.4 and STO-18. Figures 2 and 3 present the temperature dependence of the optical retardation $\delta\Psi(T)$ for STO, SCT-1.4 and STO-18 single crystals. Here, nominally pure STO was used as a reference material to evaluate the "regular" behavior of the $\delta\Psi_{a,c}^0(T)$-dependence of this material (see examples - Curve 2, Fig. 2 and Curve 3, Fig. 3), which is independent of the spontaneous polarization by the methods described in [47, 48]. Arrows mark the positions of the cubic-tetragonal PT ($T_a$) and the PT to the low-T polar state for STO-18 ($T_c = 24$ K) and SCT-1.4 ($T_c = 29$ K).

The difference between curve 1 (obtained experimentally) and curve 2 (Fig. 2) and curve 4 (obtained experimentally) and curve 3 (Fig. 3) can be attributed to the spontaneous polar contribution to the refraction of light $\delta\Psi_a^s(T)$ in SCT-1.4 and STO-18. The same procedure was



used for $\delta \Psi_c^s(T)$. Figures 4 and 5 show the spontaneous contributions to $\delta \Psi_i^s(T)$ and the morphic spontaneous polar birefringence for SCT-1.4 and STO-18. In the Fig. 6 and 7 dielectric hysteresis loops for STO-1.4 at different temperatures and at different electric fields are shown. It is seen from Fig. 7, that there are saturation of loops.

### 3.1. Generalized expressions of spontaneous polar contributions to the refractive index and relative optical retardation including fluctuations of the order parameter (short range order).

Within the indicatrix description of crystal optics [58], the variations of the principal dielectric impermeability components $\varepsilon_{ii}^{-1}$ due to the polarization (the polar contribution) [38] are given by

$$\delta \varepsilon_{ii}^{-1} = \sum_{j=1}^{3} g_{ij}^* P_j^2 \qquad (7)$$

where $P_j$ - the spontaneous polarization components along possible polar directions, $g_{i3}^*$ are the quadratic EO coefficients in the polar phase. For principal values of refractive index $\varepsilon_{ii} = n_i^2$, than for small variation of refractive index $\delta n_i^s$ due to the polarization (the polar contribution to the refractive index)

$$\delta n_i^s = -\left[\frac{n_i^3}{2}\right]\delta(\varepsilon_{ii}^{-1})^s = -\sum_{j=1}^{3}\left[\frac{n_i^3}{2}\right]g_{ij}^* P_j^2 \qquad (8)$$

Keeping in mind that we measure time- and space-averaged values $\langle P_j^2 \rangle$, $j=1,2,$ and 3, taking into account the spatial fluctuations of the polarization and that for absolute values $n_i \cong n$ from Ex. (8) we obtain

$$\delta n_i^s = -\sum_{j=1}^{3}\left[\frac{n^3}{2}\right]g_{ij}^*\langle(P_{sj} \pm P_{shj})^2\rangle \qquad (9)$$

In Ex. (9) $P_{sj}$ are long scale (long range order) spontaneous polarization components and $P_{shj}$ are polarization components which appears only due to strong spatial fluctuations ("local" or "short range" ordering). In our notations $P_{sj}$ and $P_{shj}$ are the absolute values. Average values $\langle \pm 2P_{sj}P_{shj}\rangle = 0$ due to the random distribution of order parameter fluctuations components $P_{shj}$ with opposite sign. Than we obtain generalized expressions of spontaneous polar contributions including fluctuations (short range order) to the refractive index



$$\delta n_i^s = -\sum_{j=1}^{3}\left[\frac{n^3}{2}\right]g_{ij}^*\left(\left\langle P_{sj}^2\right\rangle + \left\langle P_{shj}^2\right\rangle\right) \tag{10}$$

Ex. (10) is a general formal expression. Really axes along the principal axes of the optical indicatrix have to be used and the long scale (long range order) spontaneous polarization should be lain along one of the directions for single domain samples. In order to elucidate how to use Ex. (10) for our purpose let us consider ferroelectric phase transition in $KNbO_3$ m3m $\rightarrow$ 4mm ($C_{4v}$). Spontaneous polarization lies along one of [100] directions indexed as 3. Than from (10)

$$\delta n_3^s = -\frac{n^3}{2}\left[g_{33}^*\left(\left\langle P_{s3}^2\right\rangle + \left\langle P_{sh3}^2\right\rangle\right) + g_{31}^*\left\langle P_{sh1}^2\right\rangle + g_{32}^*\left\langle P_{sh2}^2\right\rangle\right] \tag{11a}$$

$$\delta n_1^s = -\frac{n^3}{2}\left[g_{13}^*\left(\left\langle P_{s3}^2\right\rangle + \left\langle P_{sh3}^2\right\rangle\right) + g_{11}^*\left\langle P_{sh1}^2\right\rangle + g_{12}^*\left\langle P_{sh2}^2\right\rangle\right] \tag{11b}$$

$$\delta n_2^s = -\frac{n^3}{2}\left[g_{23}^*\left(\left\langle P_{s3}^2\right\rangle + \left\langle P_{sh3}^2\right\rangle\right) + g_{21}^*\left\langle P_{sh1}^2\right\rangle + g_{22}^*\left\langle P_{sh2}^2\right\rangle\right] \tag{11c}$$

In the polar phase 4mm ($C_{4v}$) EO coefficients $g_{i3}^*$ relate to the $g_{ij}$ in the cubic phase [38, 48, 54] as $g_{33}^* = g_{11}$, $g_{23}^* = g_{12}, g_{13}^* = g_{12}$ and are symmetrical on indexes. $P_{s3} = P_s$. Fluctuations in ferroelectric phase $C_{4v}$ should be isotropic as minimum along 1 and 2 axes $P_{sh1} = P_{sh2} = P_{sh}$. Above PT fluctuations should be isotropic $P_{shj} = P_{sh}$. Than below $T_c$

$$\delta n_3^s = -\frac{n^3}{2}\left[g_{11}\left(\left\langle P_s^2\right\rangle + \left\langle P_{sh3}^2\right\rangle\right) + 2g_{12}\left\langle P_{sh}^2\right\rangle\right], \ \delta n_1^s = \delta n_2^s = -\frac{n^3}{2}\left[g_{12}\left(\left\langle P_s^2\right\rangle + \left\langle P_{sh3}^2\right\rangle\right) + (g_{11} + g_{12})\left\langle P_{sh}^2\right\rangle\right] \tag{12a}$$

For isotropic fluctuations along 3 axes below $T_c$

$$\delta n_3^s = -\frac{n^3}{2}\left[g_{11}\left\langle P_s^2\right\rangle + (g_{11} + 2g_{12})\left\langle P_{sh}^2\right\rangle\right], \ \delta n_1^s = \delta n_2^s = -\frac{n^3}{2}\left[g_{12}\left\langle P_s^2\right\rangle + (g_{11} + 2g_{12})\left\langle P_{sh}^2\right\rangle\right] \tag{12b}$$

For precursor polarization above $T_c$ ($P_s = 0$)

$$\delta n_1^s = \delta n_2^s = \delta n_2^s = -\frac{n^3}{2}\left(g_{11} + 2g_{12}\right)\left\langle P_{sh}^2\right\rangle \tag{12c}$$

Expressions (12b) and (12c) for fluctuation contribution are relatively simple because of the simple relations between $g_{i3}^*$ and $g_{ij}$. For precursor polarization above $T_c$ we obtained the same expression as was used by Kleemann and others for precursor polarization in $KNbO_3$ [reference [44] expression (5)]. By the same way it is easy to obtain the expression (1) from [39] which was used by Burns G. and Dacol to evaluate $P_{sh}$ in polar nanoregions phase in relaxors. The generalized phenomenological expressions (11) of spontaneous polar contributions to refractive index including short range order are universal and allow to calculate $P_{sh}$ above and below $T_c$ from temperature variations of polar



contribution to the refraction as well as the $P_{sh}$ in polar nanoregions and glass-like systems without long rang order. Most importance of (10) is that the expressions (10) are the set of three equations. Measurements of temperature variations of three principal values of refractive index in single domain sample after extracting the polar contribution allow one in principle to obtain as the magnitude and temperature dependence of $P_s$ so $P_{sh}$, if the $P_{sh}$ are isotropic as minimum alone two axes and the direction of the polarization is known. In the case of the isotropic fluctuations alone three axes it is more than enough two equations. It is important to note for future discussion, that in the case of isotropic fluctuations as it follows from (12) morphic spontaneous polar birefringence $\Delta n_{13} = \Delta n_{23} = n_3 - n_1 = n_3 - n_2$ is zero above $T_c$, below $T_c$ the contribution of fluctuations to $\Delta n_{13} = \Delta n_{23}$ is absent and for strong spatial fluctuations the magnitude of polar contribution to refractive index $\delta n_i^s$ should be significantly higher than magnitude of morphic birefringence.

Following logic of the section 2.2. it is easy to write expressions similar to (10) for the spontaneous polar contribution to the optical retardation, where the "EO coefficients" are given below and later correspond to these effective coefficients like in section 2.2.. Taking into account a new meaning of the EO coefficients $g_{ij}{}^*$ and , we may to rewrite (10) as

$$\delta \Psi_i^s = -\sum_{j=1}^{3} \left[ \frac{n^3}{2} \right] g_{ij}^* \left( \left\langle P_{sj}^2 \right\rangle + \left\langle P_{shj}^2 \right\rangle \right) \tag{13}$$

### 3.2. The spontaneous polar contributions to the refraction of light for STO-18 and SCT-1.4

Let us consider possible manifestations of spontaneous polarization contribution, including fluctuations (short range order) into refraction of light, using expression (13), data of the Table 1 and numerical values $g_{ij}$. In the single domain samples of SCT and STO-18, the polar axis can be aligned along the axis $[110]_a$ or $[1\bar{1}0]_b$ (i.e., corresponding to $C_{2v}$ symmetry) [35, 37, 50, 51, 56]. The spontaneous polar contribution to the relative optical retardation $\delta \psi_i^s$ is related to $P_{sj}$ as follows from formula (13), where $P_{sj}$ corresponds to spontaneous polarization $P_{sa}$ along the $[110]_a$ or $P_{sb}$ along the $[1\bar{1}0]_b$ axis, (these axes refers to the orientation of the polarization along $a$ or $b$ in the single domain crystal). Index $j$ denote as $3$ for spontaneous polarization $P_{sa}$ along the $[110]_a$ or $P_{sb}$ along the $[1\bar{1}0]_b$ axis. The $g_{ij}{}^*$ coefficients are related to those $g_{ij}$ in the cubic $O_h^1$ phase at RT, and are shown in the Table 1. Here and in the next sections, $g_{ij}{}^*$ and $g_{ij}$ denote the coefficients of the relative optical retardation (effective EO coefficients), as previously mentioned. The $g_{ij}$ numerical values are given above. Below using expressions (13) we denote index $j$ as $3$ for spontaneous



polarization $P_{sj} = P_{s3} = P_{sa}$ along the $[110]_a$ or $P_{sj} = P_{s3} = P_{sb}$ along the $[1\overline{1}0]_b$ axis, than $g_{ij}{}^*$ denote as $g_{i3}{}^*$, $-(n^3/2)g_{i3}{}^* = A_{i3}$, Consider set of expressions (13) for the possible cases of polarization orientations 1). $P_{s3}$ along the $[110]_a$ $P_{s3} = P_{sa}$ 2) $P_{s3}$ along the $[1\overline{1}0]_b$ $P_{s3} = P_{sb}$. We also denote in the expressions of the next section $\langle P_{sha}^2 \rangle = P_{sha}^2$, $\langle P_{shb}^2 \rangle = P_{shb}^2$, $\langle P_{sa}^2 \rangle = P_{sa}^2$, $\langle P_{sb}^2 \rangle = P_{sb}^2$. Taking into account Table 1 and our notations $A_{13} = -(n^3/2)(g_{12}) = -0.27$ [m$^4$/C$^2$], $A_{23} = -(n^3/4)(g_{11}+g_{12}-g_{44}) = -0.31$ [m$^4$/C$^2$], $A_{33} = -(n^3/4)(g_{11}+g_{12}+g_{44}) = -0.92$ [m$^4$/C$^2$]. From the symmetry consideration for $C_{2v}$ $A_{13} = A_{12}$, $A_{23} = A_{32}$.

### 3.2.1. The spontaneous polar contributions to the relative optical retardation and birefringence, including fluctuations (*spontaneous polarization along* $[110]_a$ *(nominated as $P_{sa}$)*).

Using expressions (13), one can write expressions for the spontaneous polar contribution to the optical retardation $\delta\psi^s{}_i$, to the principal birefringence and morphic spontaneous polar birefringence in view of contribution of fluctuations $P_{sha}$, $P_{shb}$ along $a$ and $b$ axes respectively and spontaneous polarization $P_{sa}$ along $[110]_a$.

$$\delta\psi^s{}_a = A_{33}\, P_{sa}{}^2 + A_{33}\, P^2{}_{sha} + A_{23}\, P^2{}_{shb} = A_{33}\,(P^2{}_{sa} + P^2{}_{sha}) + A_{23}\, P^2{}_{shb}$$

$$\delta\psi^s{}_b = A_{23}\, P_{sa}{}^2 + A_{23}\, P^2{}_{sha} + A_{33}\, P^2{}_{shb} = A_{23}\,(P^2{}_{sa} + P^2{}_{sha}) + A_{33}\, P^2{}_{shb} \qquad (14).$$

$$\delta\psi^s{}_c = A_{13}\, P_{sa}{}^2 + A_{13}\, P^2{}_{sha} + A_{13}\, P^2{}_{shb} = A_{13}\,(P^2{}_{sa} + P^2{}_{sha} + P^2{}_{shb})$$

For morphic spontaneous polar birefringence $\Delta n_{ab} = n_b - n_a$

$$\Delta n_{ab} = \delta n^s{}_b - \delta n^s{}_a = \delta\psi^s{}_b - \delta\psi^s{}_a = (A_{23} - A_{33})\, P^2{}_{sa} + (A_{23} - A_{33})\, P^2{}_{sha} - (A_{23} - A_{33})\, P^2{}_{shb} \qquad (15)$$

From our measurements in SCT-1.4 morphic birefringence is absent above $T_c$ (Fig.4) and the same in STO-18 from ref. [37]. As it follows from expression (15) in this case should be $P^2{}_{sha} = P^2{}_{shb}$. So it is reasonable to assume that fluctuations along $[110]_a$ and $[1\overline{1}0]_b$ are isotropic $P^2{}_{sha} = P^2{}_{shb}$. At $P^2{}_{sh} = P^2{}_{sha} = P^2{}_{shb}$ expressions (14) transform as

$$\delta\psi^s{}_a = A_{33}\, P_{sa}{}^2 + (A_{33} + A_{23})\, P^2{}_{sh} = [-0.92\, P_{sa}{}^2 - 1.23\, P^2{}_{sh}] \qquad (16)$$

$$\delta\psi^s{}_b = A_{23}\, P_{sa}{}^2 + (A_{23} + A_{33})\, P^2{}_{sh} = [-0.31\, P_{sa}{}^2 - 1.23\, P^2{}_{sh}] \qquad (17)$$

$$\delta\psi^s{}_c = A_{13}\, P_{sa}{}^2 + 2A_{13}\, P^2{}_{sh} = [-0.27\, P_{sa}{}^2 - 0.54\, P^2{}_{sh}] \qquad (18)$$

The spontaneous polar contribution to the principal birefringence is given as follows:

$$\Delta n^s{}_{ac} = \delta\psi^s{}_c - \delta\psi^s{}_a = (A_{13} - A_{33})\, P_{sa}{}^2 + (2A_{13} - A_{33} - A_{23})\, P^2{}_{sh} = [0.65\, P_{sa}{}^2 - 0.69\, P^2{}_{sh}]$$

$$(19)$$

The morphic spontaneous polar birefringence $\Delta n_{ab} = n_b - n_a$ is given as follows:



$$\Delta n_{ab} = \delta n^s_b - \delta n^s_a = \delta \psi^s_b - \delta \psi^s_a = (A_{23} - A_{33})\, P_{sa}^{2} = [+\, 0.61\, P_{sa}^{2}] \tag{20}$$

### 3.2.2. The spontaneous polar contributions to the relative optical retardation and birefringence, including fluctuations (*spontaneous polarization along* $[1\,\overline{1}\,0]_b$ *(nominated as $P_{sb}$).*

Following similar logic for $P_{sb}$ along $[1\,\overline{1}\,0]_b$ from expression for morphic birefringence one can assume isotropic fluctuations along $[110]_a$ and $[1\,\overline{1}\,0]_b$ $P^2_{sha} = P^2_{shb}$. At $P^2_{sh} = P^2_{sha} = P^2_{shb}$ numerical expressions for this case can be written as

$$\delta \psi^s_a = A_{23}\, P_{sb}^{2} + (A_{23} + A_{33})\, P^2_{sh} = [-\, 0.31\, P_{sb}^{2} - 1.23\, P^2_{sh}] \tag{21}$$

$$\delta \psi^s_b = A_{33}\, P_{sb}^{2} + (A_{33} + A_{23})\, P^2_{sh} = [-\, 0.92\, P_{sb}^{2} - 1.23\, P^2_{sh}] \tag{22}$$

$$\delta \psi^s_c = A_{13}\, P_{sb}^{2} + 2A_{13}\, P^2_{sh} = [-\, 0.27\, P_{sb}^{2} - 0.54\, P^2_{sh}] \tag{23}$$

The spontaneous polar contribution to the principal birefringence is given as follows:

$$\Delta n^s_{ac} = \delta \psi^s_c - \delta \psi^s_a = (A_{13} - A_{23})\, P_{sb}^{2} + (2A_{13} - A_{33} - A_{23})\, P^2_{sh} = [0.04\, P_{sb}^{2} - 0.69\, P^2_{sh}] \tag{24}$$

The morphic spontaneous polar birefringence $\Delta n_{ab} = n_b - n_a$ is given as follows:

$$\Delta n_{ab} = \delta n^s_b - \delta n^s_a = \delta \psi^s_b - \delta \psi^s_a = (A_{33} - A_{23})\, P_{sb}^{2} = [-\, 0.61\, P_{sb}^{2}] \tag{25}$$

### 3.3. Calculations of the long ($P_s$) and short range ($P_{sh}$) polarization.

Let us consider possible ways for calculations of $P_s$ and $P_{sh}$ from our experimental data.

1). From expressions (16,18) and (21,23) it is seen that measurements of temperature variations of $\delta \psi^s_c$ and $\delta \psi^s_a$, after extracting the polar contribution allow us to obtain magnitudes and temperature dependences of both values $P_s$ and $P_{sh}$ without any additional data, if the direction of the spontaneous polarization is known ($P_{sb}$ or $P_{sa}$).

2). From expressions (16,18, 20) and (21,23,25) it is clear that temperature variations of $\delta \psi^s_c$ or $\delta \psi^s_a$ data and morphic birefringence data $\Delta n_{ab}$ also allow us to obtain $P_s$ and $P_{sh}$, if the direction of the spontaneous polarization is known ($P_{sb}$ or $P_{sa}$). There is also additional reason to calculate $P_s$ from $\Delta n_{ab}$ according to expressions (20) or (25). The accuracy in $\delta \psi(T)$ prolonged temperature measurements of about $10^{-5}$ per hour. Thus the accuracy of $\delta \psi(T)$ measurements in the temperature range 13-300 K decrease up to 3-5 $10^{-5}$. The accuracy of $\Delta n_{ab}$ measurements is about $10^{-6}$. In this case one will have more precision calculations of $P_s$ and correspondingly $P_{sh}$ from equations above.

3) The data of temperature variations both of $\delta \psi^s_c$ and $\delta \psi^s_a$ and morphic birefringence $\Delta n_{ab}$ allow from expressions (16, 18, 20) and (21,23,25) to obtain $P_s$ and $P_{sh}$ and to verify the direction of the spontaneous polarization ($P_{sb}$ or $P_{sa}$) by comparing the calculations from two set of results: a) temperature variations of $\delta\psi^s_a$ and morphic birefringence $\Delta n_{ab}$ b) temperature variations of $\delta\psi^s_c$ and morphic birefringence $\Delta n_{ab}$.

It is clear that the third way give us more opportunities, including the selection of the real direction of the spontaneous polarization, and more precision calculations. Measurements of the spontaneous polarization $P_s$ from dielectric hysteresis loops in these cases really have meaning as a checkup the $P_s$ data obtained from $\Delta n_{ab}$.

We determined $P_s$ magnitude from morphic birefringence $\Delta n_{ab}$ according to expressions (20,25) and than after substituting $P_s = P_{sMB}$ into (16,18 and 21,23) $P_{sh}$ were calculated from $\delta\psi^s_c$ or $\delta\psi^s_a$ according to two sets of expressions (16,18) and (21,23). For this purposes we used our morphic birefringence data for SCT-1.4, morphic birefringence data for STO-18 reported earlier in [37] with the participation of one of our co-author M. Itoh and our specific optical retardation data for SCT-1.4, and STO-18. We have performed such calculations for SCT-1.4 and STO-18 for the both possible orientations of spontaneous polarization $P_s$ (along $[110]_a$ and $[1\bar{1}0]_b$). As result, we succeed to calculate $P_{sh}$ from independent expression systems (16, 20), (18, 20) for $P_s$ along $[110]_a$ and (21, 25), (23, 25) for $P_s$ along $[1\bar{1}0]_b$. It was turned out to be that only the case when $P_s$ is oriented along $[1\bar{1}0]_b$.agree well with calculation of $P_{sh}$ from $\delta\psi^s_a$ and $\delta\psi^s_c$. Such orientation of $P_s$ agrees with suggestions [35, 37, 50]. All our calculations results following below consider situation when $P_s$ is oriented along $[1\bar{1}0]_b$. Figures 8 and 9 present calculation results for SCT-1.4 и STO-18. Our notations $P_s$ and $P_{sh}$ correspond to the absolute value of polarizations. Figures 8 and 9 present also spontaneous polarization obtained from dielectric hysteresis loops measurements $P_s = P_{sLOOPS}$ obtained from our measurements for SCT-1.4 and evaluated from [10,18] for STO-18. Although $P_{sMB}$ and polarization was taken from measurements of hysteresis loops $P_{sLOOPS}$ coincides at low temperatures or are close, $P_{sMB}$ reflects namely spontaneous polarization without electric field effect (in comparison with $P_{sLOOPS}$ at hysteresis loops measurements).

Dependences for $P_{sh}(T)$, presented in the Fig.8 and Fig.9 are given for calculations using Eq. (21) from $\delta\psi^s_a$. As result magnitude of $P_{sh}$, calculated from $\delta\psi^s_c$ coincides with $P_{sh}$ is given in the Figs. 8, 9 with the accuracy of experimental data used.

From Eq. (24) it is seen that $P_{sh}(T)$ can be calculated from measurements of spontaneous polar contribution to the principal birefringence, if magnitude of $P_s$. is known. We performed such calculations by our method for $Sr_{1-x}Ca_xTiO_3$ (x = 0.007). We analyzed the data from the spontaneous polar contribution to the principal birefringence [35, 50] and the morphic polar birefringence [35, 50], as well as the data from our hysteresis loops measurements (to be published). The magnitudes $P_s$ and $P_{sh}$ evaluated for $Sr_{1-x}Ca_xTiO_3$ (x = 0.007) are shown in the Table 2 at low temperatures. It was observed that at low temperatures in $Sr_{1-x}Ca_xTiO_3$ (x = 0.007) $P_{sMB} \approx P_{sLOOPS}$. The Table 2 presents also magnitudes of $P_s$ and $P_{sh}$ at low temperatures for SCT-1.4 and STO-18. For STO-18 we gave magnitudes $P_s = P_{sMB}$. Values of $P_{sLOOPS}$, estimated from literature data [10, 18, 19] are turned out to be rather close.

Table 2. Characteristics of induced polar states in quantum paraelectrics. $P_s$ и $P_{sh}$ at low temperatures.

| Systems | $P_{sh}$ C/m$^2$ | $P_s = P_{sMB} \approx P_{sLOOPS}$ C/m$^2$ |
|---|---|---|
| $Sr_{1-x}Ca_xTiO_3$ (x = 0.007) | $2.5\ 10^{-2}$  (5.5 K) | $0.7\ 10^{-2}$  (5.5 K) |
| $Sr_{1-x}Ca_xTiO_3$ (x = 0.014) | $2.7\ 10^{-2}$  (14 K) | $1.8\ 10^{-2}$  (14 K) |
| $SrTi(^{16}O_{1-x}{}^{18}O_x)_3$ (x = 0.97) | $3.0\ 10^{-2}$  (13 K) | $1.9\ 10^{-2}$  (13 K) |

Data of the Table 2 show that in $Sr_{1-x}Ca_xTiO_3$ at the approaching to critical concentration $x_c = 0.0018$ from x = 0.014 to x =0.007 spontaneous polarization $P_s$ decreases sufficiently whereas $P_{sh}$ does not change significantly. It is reflected the role of impurities or oxygen isotope exchange inducing long range polar order in $SrTiO_3$ as defects leading to spatial fluctuations in the order parameter and short range ordering phenomena. Our results does not allow to formulate microscopic model of the polar phase formation in the objects under study. We only can assume the following scenario. Above $T_C$ spatial fluctuations of the polarization isotropic in *ab* plane ($P^2_{sh} = P^2_{a\ sh} = P^2_{b\ sh}$) appear. At cooldown magnitude of $P_{sh}$ increases and at $T_C$ a percolation-type phase transition takes place in which correlation of fluctuations together and formation of the a long range order with $P_s$ coexist. Then, at further cooling fluctuations continue to increase as well as $P_s$. In a certain sense this simple scenario is close to qualitative suggestions had been offered in [5, 35, 57] for SCT (0.007) (ordering at two length scales: local ordering associated with individual nanodomains leading to the relaxor character of the response, and ordering on a longer length scale). Similar model of various versions had been discussed at various times [4-5]. However, quantitative comparable characteristics of short- and long range contributions into polar phase formation below Tc have not been presented. It is done in our present study for the first time. Naturally, spatial polar fluctuations can be inhomogeneous ones along the crystals. Therefore magnitude of $P_{sh}$, which we have obtained, is a some mean (average) value due to time and space-averaged of the square of the polarization.

## 4. CONCLUSIONS

In this report we present the investigations of short and long range impurity-induced polar ordering in $Sr_{1-x}Ca_xTiO_3$ with x = 0.014 (SCT-1.4) and $SrTi(^{16}O_{0.03}{}^{18}O_{0.97})_3$ (STO-18) single crystals. The temperature dependence of the principal values of the refraction (specific optical retardation) of light in STO-18 and SCT-1.4 has been measured and the magnitudes and temperature variations of polar contributions to the refraction were derived. It is shown that interferometric measurements of specific optical retardation temperature and electric field dependences are sufficient to obtain root mean square of the polarization $\left\langle P^2 \right\rangle^{1/2}$ without additional temperature dilatation measurements. A method of quantitative calculations from polar contributions to the refraction of light on the polarization $P_{sh} = \left\langle P_{sh}^2 \right\rangle^{1/2}$ which appears only due to strong spatial fluctuations ("local" or "short-range" ordering) was developed for the systems with the coexistence of long and short range polar ordering within the framework of phenomenological indicatrix perturbation theory from a generalized expression of polarization variation (including short-range ordering) of the dielectric impermeability components. The temperature dependences of the morphic birefringence, dielectric hysteresis loops for SCT-1.4 and electrooptic (EO) coefficients for SCT-1.4 and STO-18 were measured too in order to calculate more carefully and strictly separately $P_{sh}$ and $P_s$ below $T_C$. As a result we determined for SCT-1.4 and STO-18 the magnitude and temperature dependence of $P_{sh}$ polarization which appears only due to strong spatial fluctuations ("local" or "short range" ordering) and leads to contribution of short range order in the polar phase below and above $T_C$ as well as the magnitude and temperature dependence of long range order polarization $P_s$. These results allowed us to characterize quantitatively and to compare contributions of short ($P_{sh}$) and long ($P_s$) range ordering in the formation of impurity-induced polar phase in $SrTiO_3$. It was shown that at low temperatures in SCT-1.4 and STO-18 the magnitude of $P_{sh}$ one and a half higher than $P_s$. At low temperatures the magnitudes of $P_s$ and $P_{sh}$ were also evaluated by our method for $Sr_{1-x}Ca_xTiO_3$ (x = 0.007) from birefringence data reported in [35, 50]. In $Sr_{1-x}Ca_xTiO_3$ approaching to the critical concentration $x_c$ = 0.0018 from x = 0.014 to x =0.007 is accompanied by spontaneous polarization $P_s$ rapidly decreasing, whereas $P_{sh}$ does not change significantly. Our discussions of experimental results based on the phenomenological approach and does not allow in a unique manner to formulate microscopic model of the impurity-induced polar phase in $SrTiO_3$. Nevertheless the obtained results showed quantitatively important role of impurities (or oxygen isotope exchange) that induced both long-range polar order while acting as defects that induce spatial fluctuations in the order parameter and short range ordering phenomena.

We would like to note that the method developed for quantitative calculations of short ($P_{sh}$) and long ($P_s$) range contributions to the formation of the polar phase was applied for STC and STO18 single crystals with coexistence of long and short range ordering. Really experimental results and calculations $P_{sh}$ and $P_s$ in STC and STO18 should be consider as a convenient experimental example of our approach. We would like to underline that the method developed can be used for other systems with coexistence of long and short polar ordering below $T_C$.

## ACKNOWLEDGMENTS

We thank J. G. Bednorz and W. Kleemann for kindly providing us with the STO and SCT-1.4 crystals.

REFERENCES

[1] B. E. Vugmeister and M. D. Glinchuk, Rev. Mod. Phys. **62**, 993 (1990).

[2] U. T. Hoechli, K. Knorr and Loidl, Adv. in Phys. **39**, 405 (1990).

[3] M. D. Glinchuk and I. P. Bykov, Phase Transitions **40**, 1 (1992).

[4] W. Kleemann, Int. J. Mod. Phys. B **7**, 2469 (1993).

[5] G. A. Samara, J. Phys.: Condens. Mat. **15**, R367 (2003).

[6] S. A. Prosandeev, V. A. Trepakov, JETP **94**, 419 (2002).

[7] O. E. Kvyatkovskii, Phys. Sol. State **43**, 1401 (2001).

[8] O. E. Kvyatkovskii, Phys. Sol. State **44**, 1135 (2002).

[9] O. E. Kvyatkovskii, Ferroelectrics **314**, 143 (2005).

[10] M. Itoh, R. Wang, Y. Inaguma, T. Yamaguchi, Y-J Shan and T. Nakamura, Phys. Rev. Lett. **82**, 3540 (1999).

[11] V. S. Vikhin, P. A. Markovin, V. V. Lemanov, W. Kleemann, J. Korean Phys. Soc. **32**, S853 (1998).

[12] O. E. Kvyatkovskii, Solid State. Comm. **117**, 455 (2001).

[13] A. Bussmann-Holder, H. Büttner, A. R. Bishop, J. Phys.: Cond. Mat. **21**, L115 (2000).

[14] A. Bussmann-Holder, A. R. Bishop, Eur. Phys. J. **53**, 279 (2006).

[15] Y. Minaki, M. Kobayasi, Y. Tsujimi, T. Yagi, N. Nakanishi, R. Wang, M. Itoh, J. Korean Phys. Soc. **42**, S1290 (2003).

[16] M. Takesada, M. Itoh, T. Yagi, Phys. Rev. Lett. **96**, 227602 (2006).

[17] E. L. Venturini, G. Samara, M. Itoh, R. Wang, Phys. Rev. B.**69**, 184105 (2004).

[18] C. Filipic and A. Levstik, Phys. Rev. B **73**, 092104 (2006).

[19] C. Filipic and A. Levstik, Phys. Rev. B **75**, 027102 (2007).

[20] R. Blinc, B. Zalar, V. V. Laguta, M. Itoh, Phys. Rev. Lett. **94**, 147601 (2005).

[21] B. Zalar, A. Lebar, J. Seliger, R. Blinz, V. V. Laguta, M. Itoh, Phys. Rev. B **71**, 064107 (2005).

[22] V. V. Laguta, R. Blinz, M. Itoh, J. Seliger, B. Zalar, Phys. Rev. B. **72**, 214117 (2005).

[23] R. Wang and M. Itoh, Phys. Rev. B **64**, 174104 (2001).

[24] H. Taniguchi, T. Yagi, M. Takesada, M. Itoh, Phys. Rev. B. **72**, 064111 (2005).

[25] R. Wang, M. Itoh, Phys. Rev. B. **62**, R731 (2000).

[26] R. Wang, M. Itoh, Phys. Rev. B. **64**, 174104 (2001).

[27] W. Kleemann, J. Dec, R. Wang, M. Itoh, Phys. Rev. B **67**, 092107 (2003).

[28] J. Dec, W. Kleemann, M. Itoh, Phys. Rev. B **71**, 144113 (2005).

[29] J. Dec, W. Kleemann, M. Itoh, Ferroelectrics **316**, 59 (2005).

[30] Y. Uesu, R. Nakai, J-M Kiat, C. Menoret, M. Itoh, T. Kyomen, J. Phys. Soc. Japan.**73**, 1139 (2004).

[31] H. Taniguchi, M. Itoh, T. Yagi, Phys. Rev. Lett. **99**, 017602 (2007).

[32] S. K. Mishra, R. Ranjan, D. Pandey, H. T. Stokes, J. Phys.: Condens. Mat. **18**, 1855 (2006).

[33] S. K. Mishra, R. Ranjan, D. Pandey, P. Ranson, R. Oillon, J.-P. Pinan-Lucarre, P. Puzan, J. Phys.: Condens. Mat. **18**, 1899 (2006).

[34] V. A. Trepakov, M. E. Savinov, S. A. Prosandeev, S. Kapphan, L. Jastrabik, L. A. Boatner, Phys. Stat. Sol. (c) **2**, 145 (2005).

[35] W. Kleemann, U. Bianchi, A. Buergel, M. Prasse, J. Dec, Phase Transitions. **55**, 57 (1995).

[36] W. Kleemann, F. J. Schäfer, D. Rytz, Phys. Rev. Lett. **54**, 2038 (1985).

[37] T. Azuma, K. Iio, K. Yamanaka, T. Kyomen, R. Wang, M. Itoh, Ferroelectrics **304**, 77 (2004).

[38] M. Di Domeniko, S. H. Wemple, J. Appl. Phys. **40**, 720 (1969).

[39] G. Burns and F. H. Dacol, Solid State Commun. **48**, 853 (1983).

[40] O.Yu. Korshunov, P. A. Markovin, and R. V. Pisarev, Sov. Phys. Sol. State, **25**, 1228 (1983).

[41] O.Yu. Korshunov, P. A. Markovin, R. V. Pisarev, Ferroelectrics Lett. **13**, 137 (1992).

[42] O.Yu. Korshunov, P. A. Markovin, R. V. Pisarev, L. M. Sapoznikova, Ferroelectrics **90**, 151 (1989).

[43] G. Burns and F. H. Dacol, Ferroelectrics **37**, 661 (1981).

[44] W. Kleemann, F. J. Schafer, and M. D. Fontana, Phys. Rev. B **30**, 1148 (1984).

[45] W. Kleemann, F. J. Schafer and D. Rytz, Phys. Rev. B **34**, 7873 (1986).

[46] R. V. Pisarev, P. A. Markovin, B. N. Shermatov, V. I. Voronkova, V. K. Yanovskii, Ferroelectrics **96**, 181 (1989).

[47] P. A. Markovin, W. Kleemann, R. Linder, V. V. Lemanov, O. Yu. Korshunov, P. P. Syrnikov, J. Phys.: Condens. Mat. **8**, 2377 (1996).

[48] M. E. Guzhva, V. V. Lemanov, P. P. Markovin, W. Kleemann, Phys. Sol. State **39**,  618 (1997).

[49] K. Yamanaka, R. Wang., M. Itoh, K. Iio, J. Phys. Soc. Japan **70**, 3213 (2001).

[50] U. Bianchi, PhD Thesis, Gerhard-Mercator-Universität-Gesamthochschule-Duisburg, Duisburg (1996), 159 p.

[51] J. G. Bednorz and K. A. Miiller, Phys. Rev. Lett. **52**, 2289 (1984).

[52] P. A. Markovin, R. V. Pisarev, Sov. Phys. JETP **50**, 1190 (1979).

[53] R. V. Pisarev, B. B. Krichevtzov, P. A. Markovin, O. Yu. Korshunov, J. F. Scott, Phys. Rev. B 28, 2677 (1983).

[54] Y. Fuji, T. Sakudo, J. Appl. Phys. **41**, 4118 (1970).

[55] J. E. Geusiz, S. K. Kuttz, L. G. Van Uitert, S. H. Wemple, Appl. Phys. Lett. **4**, 141 (1964).

[56] T. Shigenari, K. Abe, T. Takeoto, O. Sanaka, T. Akaike Y. Sakai, R. Wang, M. Itoh, Phys. Rev. B **74**, 174121 (2006).

[57] W. Kleemann, A. Albertini, M. Kuss and R. Lindner, Ferroelectrics **203**, 57 (1997).

[58] J. F. Nye, Physical Properties of Crystals (Oxford University Press, Oxford 1979). 1960, p. 245.

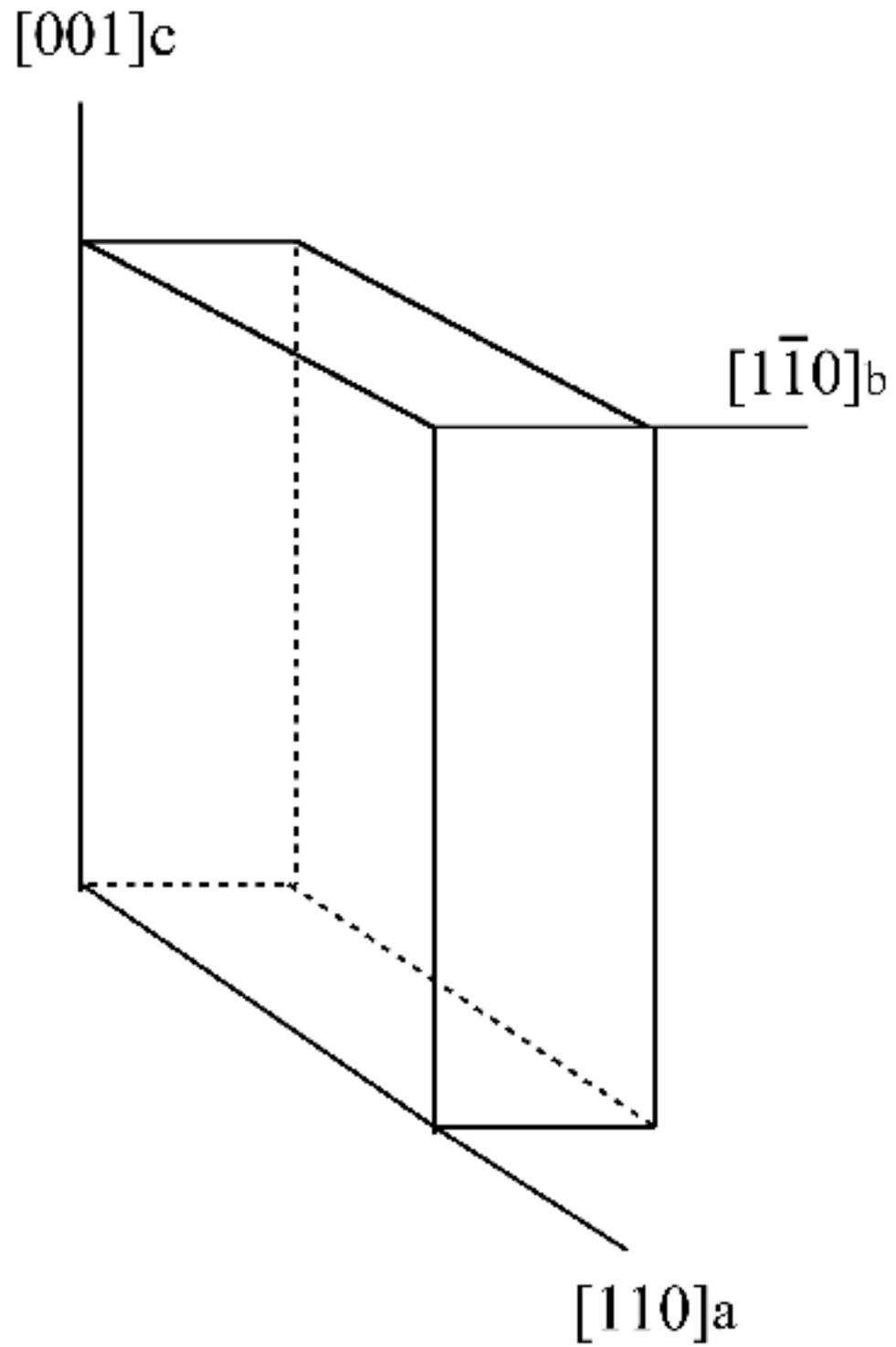

**Figure 1.** Axes designation for STO-1.4 and STO-18 specimens

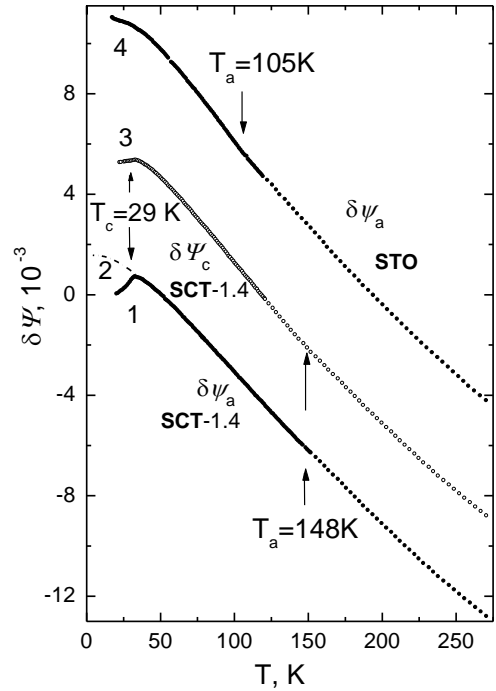

**Figure 2.** Relative optical retardation $\delta\Psi(T)$ for STO (4) and STO-1.4 (1, 3) single crystals, and the extrapolation (Curve 2) for the regular contribution $\delta\Psi_a^0(T)$. Arrows show phase transition points.

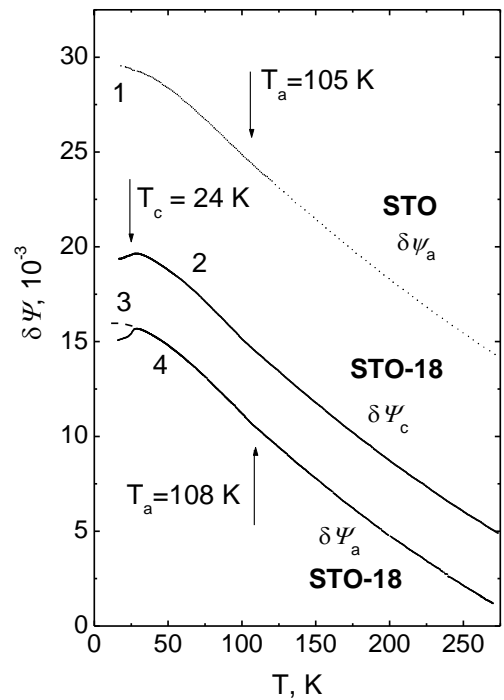

**Figure 3.** Relative optical retardation $\delta\Psi(T)$ for STO (1) and STO-18 (2,4) single crystals and extrapolation of the temperature dependence of the regular contribution $\delta\Psi_a^0(T)$ - (Curve 3).

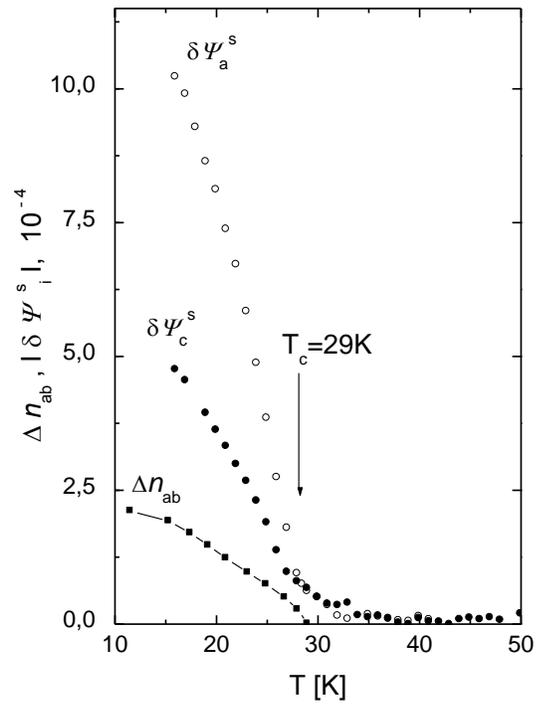

**Figure 4.** The absolute values of spontaneous polar contributions to the $\delta\Psi_i$ for STO-1.4 and the morphic spontaneous polar birefringence $\Delta n_{ab} = n_b - n_a$.

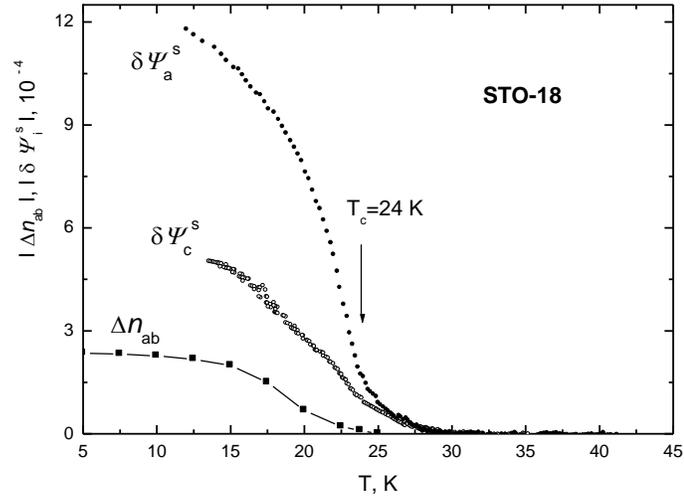

**Figure 5.** The absolute values of spontaneous polar contributions to the $\delta \Psi_i$ for STO-18 and the morphic spontaneous polar birefringence $\Delta n_{ab} = n_b - n_a$ (from Ref. [37]) for STO-18.

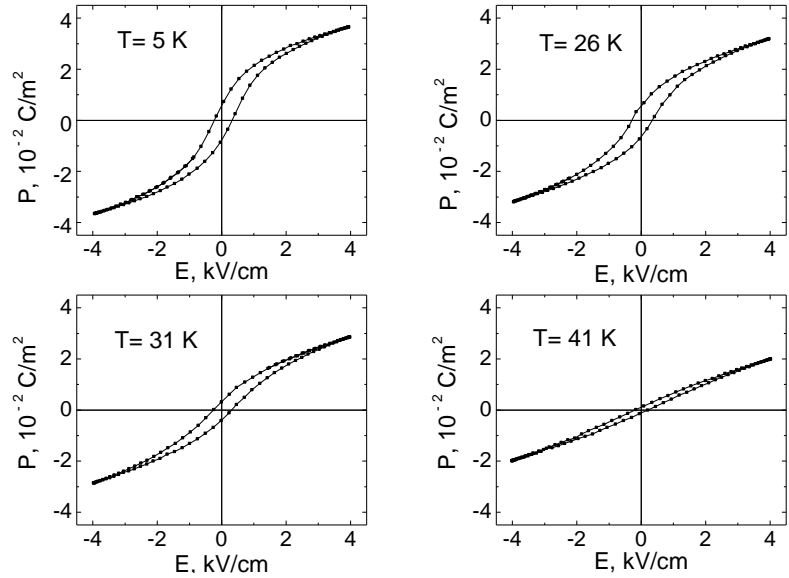

**Figure 6.** Dielectric hysteresis loops for STO-1.4 at different temperatures.

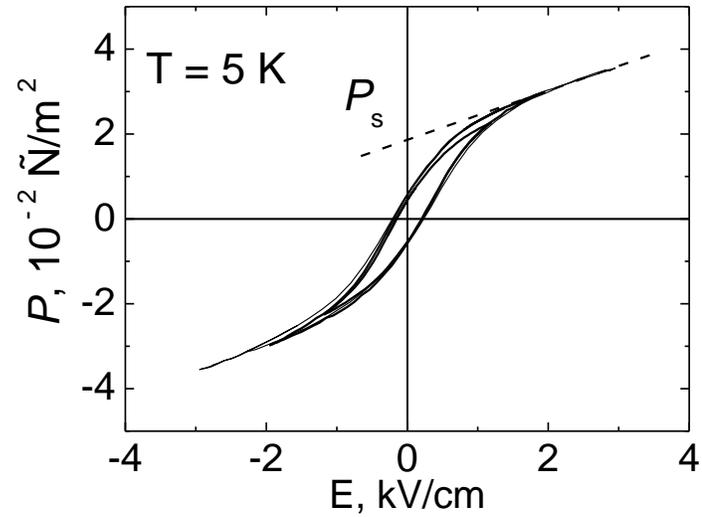

**Figure 7.** Dielectric hysteresis loops for STO-1.4 were taken at different electric fields.

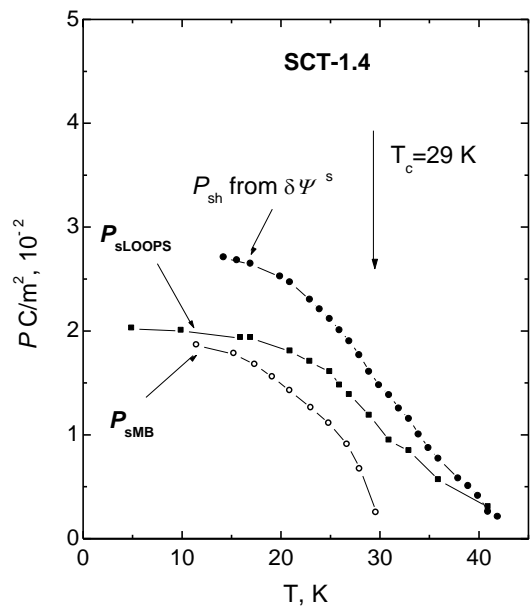

**Figure 8.** Spontaneous polarization obtained from morphic birefringence $P_{sMB}$ and from dielectric hysteresis loops measurements $P_{sLOOPS}$; $P_{sh}(T)$ dependence in SCT-1.4 is calculated using formula (21).

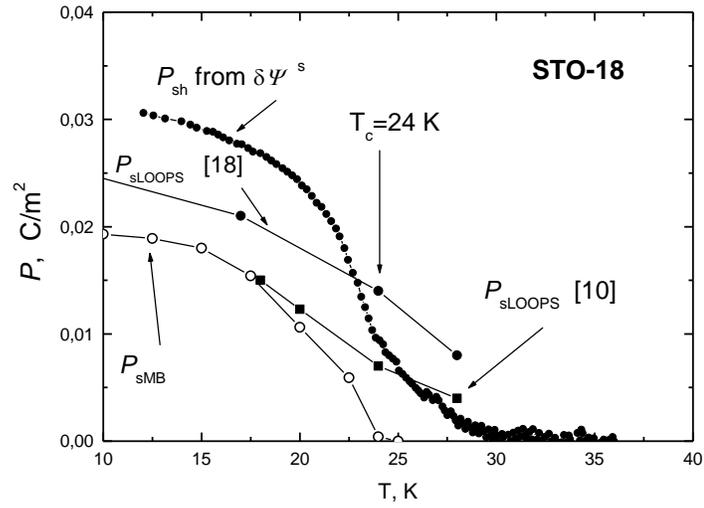

**Figure 9.** Spontaneous polarization $P_{sMB}$ obtained from morphic birefringence [37] and from dielectric hysteresis loops $P_{sLOOPS}$ [10,18]; $P_{sh}$(T) dependence in STO-18 is calculated using formula (21). Numbers in squared brackets cite references from which $P_{sLOOPS}$ magnitude was evaluated.